\documentclass[prd,twocolumn,a4paper,superscriptaddress]{revtex4}
\usepackage{graphicx,amssymb}
\begin{document}
%My commands
\newcommand{\be}{\begin{equation}}
\newcommand{\ee}{\end{equation}}
\newcommand{\bq}{\begin{eqnarray}}
\newcommand{\eq}{\end{eqnarray}}
\newcommand{\bsq}{\begin{subequations}}
\newcommand{\esq}{\end{subequations}}
\newcommand{\bc}{\begin{center}}
\newcommand{\ec}{\end{center}}
\newcommand {\R}{{\mathcal R}}
\newcommand{\al}{\alpha}
\newcommand\lsim{\mathrel{\rlap{\lower4pt\hbox{\hskip1pt$\sim$}}
    \raise1pt\hbox{$<$}}}
\newcommand\gsim{\mathrel{\rlap{\lower4pt\hbox{\hskip1pt$\sim$}}
    \raise1pt\hbox{$>$}}}

\title{Defect Junctions and Domain Wall Dynamics}
\author{P.P. Avelino}
\email[Electronic address: ]{ppavelin@fc.up.pt}
\affiliation{Centro de F\'{\i}sica do Porto, Rua do Campo Alegre 687, 4169-007 Porto, Portugal}
\affiliation{Departamento de F\'{\i}sica da Faculdade de Ci\^encias
da Universidade do Porto, Rua do Campo Alegre 687, 4169-007 Porto, Portugal}
\author{C.J.A.P. Martins}
\email[Electronic address: ]{C.J.A.P.Martins@damtp.cam.ac.uk}
\affiliation{Centro de F\'{\i}sica do Porto, Rua do Campo Alegre 687, 4169-007 Porto, Portugal}
\affiliation{Department of Applied Mathematics and Theoretical Physics,
Centre for Mathematical Sciences,\\ University of Cambridge,
Wilberforce Road, Cambridge CB3 0WA, United Kingdom}
\author{J. Menezes}
\email[Electronic address: ]{jmenezes@fc.up.pt}
\affiliation{Centro de F\'{\i}sica do Porto, Rua do Campo Alegre 687, 4169-007 Porto, Portugal}
\author{R. Menezes}
\email[Electronic address: ]{rms@fisica.ufpb.br}
\affiliation{Centro de F\'{\i}sica do Porto, Rua do Campo Alegre 687, 4169-007 Porto, Portugal}
\affiliation{Departamento de F\'\i sica, Universidade Federal da Para\'\i ba,\\
 Caixa Postal 5008, 58051-970 Jo\~ao Pessoa, Para\'\i ba, Brazil}
\author{J.C.R.E. Oliveira}
\email[Electronic address: ]{jeolivei@fc.up.pt}
\affiliation{Centro de F\'{\i}sica do Porto, Rua do Campo Alegre 687, 4169-007 Porto, Portugal}
\affiliation{Departamento de F\'{\i}sica da Faculdade de Ci\^encias
da Universidade do Porto, Rua do Campo Alegre 687, 4169-007 Porto, Portugal}

\date{25 April 2006}

\begin{abstract}
We study a number of domain wall forming models where various types of defect junctions can exist. These illustrate some of the mechanisms that will determine the evolution of defect networks with junctions. Understanding these mechanisms is vital for a proper assessment of a number of cosmological scenarios: we will focus on the issue of whether or not cosmological frustrated domain wall networks can exist at all, but our results are also relevant for the dynamics of cosmic (super)strings, where junctions are expected to be ubiquitous. We also define and discuss the properties that would make up the ideal model in terms of hypothetical frustrated wall networks, and provide an explicit construction for such a model. We carry out a number of numerical simulations of the evolution of these networks, analyze and contrast their results, and discuss their implications for our no-frustration conjecture. 
\end{abstract}
\maketitle

\section{\label{intr}Introduction}

If our current understanding of particle physics and unification scenarios is correct, defect networks must necessarily have formed at phase transitions in the early universe \cite{KIBBLE}. The type of defect that forms and its specific properties will depend on the particular details of each symmetry breaking, so a wide range of possibilities exist, with correspondingly different cosmological consequences---see \cite{VSH} for a thorough discussion of these possibilities. One example is what happens when two extended defects (cosmic strings or domain walls) interact with each other. In the simplest models they will intercommute (or exchange partners); for cosmic strings, for example, this mechanism is crucial for the existence of the well-known linear scaling solution (commonly referred to as scaling).

It is possible, however, that topological reasons will prevent intercommuting, and that the result of interactions is the formation of junctions (or bridges) between the original defects. An interesting question is then what, if any, is the effect of this mechanism on the dynamics of the defect networks. In particular, one needs to understand the detailed behavior of the defect junctions, and the role they play in the evolution and cosmological consequences of the network. These models have been much less studied \cite{RYDEN,PEN,KUBOTANI,MCRAW,NONINT,COPELAND,STEER}, though this is purely due to the inherent difficulty of doing so, and not due to any lack of physical motivation for doing it. In fact, quite the opposite is true: recent work has shown \cite{SARANGI,VILENKIN,POLCH} that cosmic (super)strings will necessarily form at the end of brane inflation, and it is believed (though not yet unambiguously proved) that such networks will have junctions. Understanding their role should therefore be a key step in studying the evolution of these networks, and in particular for determining whether or not a linear scaling solution is also an evolutionary attractor in this case: despite some recent supporting evidence \cite{NONINT,EPS,TYE,COPELAND}, this remains an open question. Even though understanding the evolution of cosmic (super)string networks is not the main motivation for the present work, it is clear that some of our results provide strong hints for what should happen there.

The present work is the second in a series whose driving goal is to understand the cosmological evolution of domain wall networks \cite{PRESS,SIMS1,SIMS2,AWALL} in sufficient detail to be able to answer the question of whether or not a frustrated defect network \cite{SOLID} can exist at all. In a previous work \cite{US} (hereafter referred to as Paper I) we have shown that the answer is trivially no for the case of cosmic strings, and provided arguments leading to the conjecture that the answer should also be no for domain walls. We have also established the simple but crucial (and yet previously neglected) result that in models with more than two vacua, if all the domain walls connecting the various vacua have equal energies then only Y-type stable junctions will form. Here we focus on the role of junctions, and we will be studying several models where different kinds of junctions can appear.

Specifically, we consider a number of realizations of the two classes of models introduced in \cite{KUBOTANI,BAZEIA} (the former has also been subsequently considered in \cite{BATTYE}). The similarities and differences between them will allow us to point out the key mechanisms at play, and further characterize the differences between models with stable Y-type junctions, models with stable X-type junctions, and models where both types can co-exist. The details of the numerical simulations that we shall present are analogous to those in \cite{PRESS,SIMS1,SIMS2,AWALL}. We then go beyond these particular models, and discuss the properties that would make up the ideal model, at least from the point of view of obtaining hypothetical frustrated networks. We provide a general and explicit construction of this ideal model, and again numerically study some examples. The pentahedral symmetry model discussed in \cite{CARTER} turns out to have some of these properties of this ideal model. In the interest of clarity, we shall start by discussing the simplest models, and gradually move on to more elaborate ones. In particular, we start by considering two-field models, then move on to three-field models, and finally discuss the ideal model and its relation with other existing models. Also in the interest of simplicity, our simulations are done in two spatial dimensions, which is sufficient to illustrate the key processes. The cosmologically more interesting (but also more complicated) case of three spatial dimensions will be discussed in subsequent work.

%%%%%%%%%%%%%%%%%%%%%
\section{\label{twofield}Two-Field Models}

We will start by considering models described by two real scalar fields, which is the minimum configuration required in order to form networks with junctions. We will discuss in some detail the Bazeia-Brito-Losano model (henceforth referred to as the BBL model) introduced in \cite{BAZEIA}. Its original purpose was as a driver for a dynamical mechanism leading to $3+1$ space-time dimensions, but from our point of view this particular motivation is not essential and we can simply take it as a toy model, which turns out to contain a simple yet rich and illuminating set of possibilities. The model has the following Lagrangian
\be
\mathcal{L} = \frac12 \sum_{i=1}^2 (\partial_\mu \phi_i \partial^\mu \phi_i)  + V(\phi_i),
\ee 
where the $\phi_i$ are real scalar fields and the potential has the form
\be
V(\phi_i) = \frac12 \sum_{i=1}^2 \left(r-\frac{\phi_i^2}r \right)^2 + \frac{\epsilon}4 \left(\phi_1^4 + \phi_2^4 - 6 \phi_1^2 \phi_2^2 + 9\right)\,.   \label{pot1}
\ee
where $r$ and $\epsilon$ are two real parameters. This potential has minima at the vertices of a square in the plane $(\phi_1, \phi_2)$, the orientation of which depends on the value of the perturbation parameter $\epsilon$ that can be assumed to vary in the range $-2\,<\,\epsilon r^2\, <\, 1$. There are a total of six independent topological sectors connecting the different minima. In the range $ -1/2\,<\,\epsilon r^2\,<\,1$  the minima are 
\be
\phi_i^2=\frac{r^2}{1-\epsilon r^2},\quad i=1,2\,,\label{solplus}
\ee
while the range $ - 2\, <\, \epsilon r^2\, <\, - 1/2$ the minima are
\be
\phi_i^2 = \frac{r^2}{1+\epsilon r^2/2}, \quad \phi_{j \neq i}^2=0\,.\label{solminus}
\ee

\begin{figure}
\begin{center}
\includegraphics*[width=6cm]{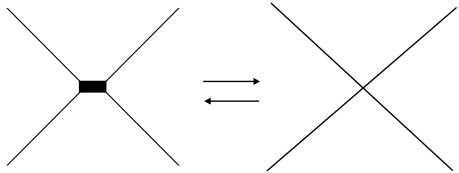}
\end{center}
\caption{Depending on whether the thick wall has a tension smaller/larger than twice that of the lower tension ones the 
formation of a Y-type/X-type junction will be favored on energetic grounds.}
\label{threefour}
\end{figure}

This model allows for Y-type and X-type junctions depending on the value of $\epsilon$, which controls the tension of 
the walls connecting each pair of vacua. Note that there are two classes of walls which we will denote simply by 
{\it{edges}} and {\it{diagonals}}. In the former the wall joins two neighboring minima in field space, and there are four such walls. In the latter the wall joins two opposite minima in field space, and there are two such walls. In Ref. \cite{BAZEIA} the authors studied only the static behavior of the model in the limit for very small $\epsilon$ and also made the choice $r=\sqrt{3/2}$. In this case, the minima of potential are in the branch given by Eq. (\ref{solplus}), that is $\phi_i^2=(2/3-\epsilon)^{-1/2}$. By integrating the $T_0^0$ component of the stress-energy tensor across the wall between each pair of vacua one obtains \cite{BAZEIA}
\be
\frac{\sigma_d}{\sigma_e} =\frac{2+3\epsilon}{1 + 21 \epsilon/8}\,,\label{sigma}
\ee
where $\sigma_e$ and $\sigma_d$ respectively denote the tension of the walls of \textit{edge} and \textit{diagonal} type. Depending on whether the diagonal walls have a tension smaller or larger than twice that of the edge ones, the formation of a Y-type or X-type junction will be favored on energetic grounds (see Fig. \ref{threefour}). In other words, in the case $\sigma_d < 2\sigma_e$ (which corresponds to $\epsilon > 0$) we expect to have only Y-type stable junctions, while for $\sigma_d > 2\sigma_e$ (which occurs for $\epsilon < 0$) we expect that only X-type stable junctions will be formed.

\begin{figure}
\begin{center}
\includegraphics*[width=6.9cm]{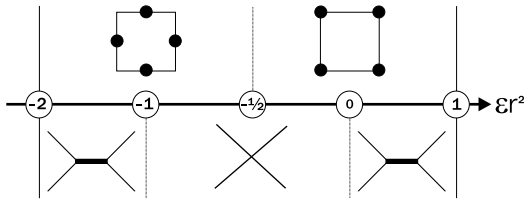}
\end{center}
\caption{Types of junctions (bottom) and configuration of the minima (top), as a function of the parameter $\epsilon r^2$ in the two-field BBL model \protect\cite{BAZEIA}.}
\label{config2d}
\end{figure}

Of course, this energetic argument is valid generically, not just in the BBL limit $\epsilon \rightarrow 0$. We have analyzed the relation between tension of edges, $\sigma_e$, and diagonals, $\sigma_d$, for the whole range $-2 <\epsilon r^2 < 1$. We found two values of $\epsilon r^2$ for which $\sigma_d=2\sigma_e$: in addition to the BBL case, $\epsilon r^2=0$, there is also $\epsilon r^2=-1$. The analysis leads us to expect that for $\epsilon r^2> 0$ and $\epsilon r^2< -1$ Y-type junctions are favored while for $-1 <\epsilon r^2< 0$ X-type junctions are preferred. These solutions are illustrated in Fig. \ref{config2d}. In passing, we note that the parameter $\epsilon$ determines not only the ratio of the energies in the two sectors, but also influences, among other things, how fast the unstable junctions will decay into the stable ones (this is an interesting and relevant question in itself, which is left for detailed study in the future).

%%%%%%%%%%%%%%%%%%%%%%%%%%%%%%%%%%%%%%%%%%%%%%%%%%%%%%%%%

%%%%%%%%%%%%%%%%%%%%%%%%%%%%%%%%%%%%%%%%%%%%%%%%%%%%%%%%%

%%%%%%%%%%%%%%%%%%%%%%%%%%%%%%%%%%%%%%%%%%%%%%%%%%%%%%%%%
\begin{figure}
\begin{center}
\includegraphics*[width=3.6cm, height=3.6cm]{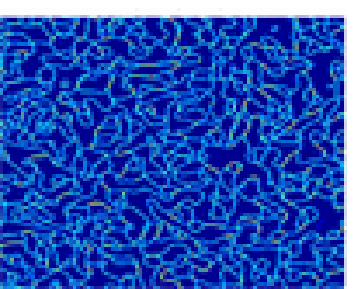}
\includegraphics*[width=3.6cm, height=3.6cm]{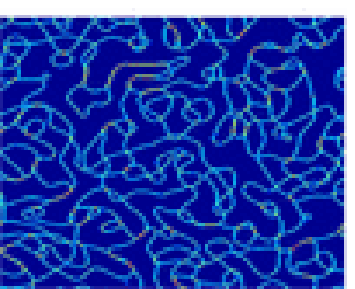}
\includegraphics*[width=3.6cm, height=3.6cm]{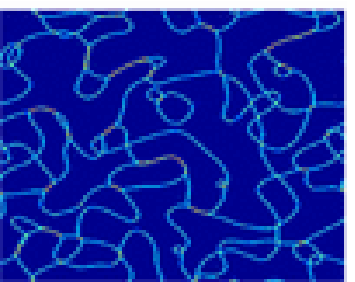}
\includegraphics*[width=3.6cm, height=3.6cm]{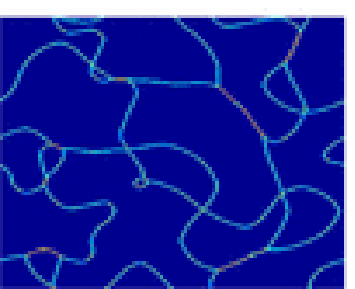}
\end{center}
\caption{The matter-era evolution of the domain wall network model of \protect\cite{BAZEIA} with potential of Eq. (\protect\ref{pot1}) and a parameter choice $r=\sqrt{3/2}$ and $\epsilon = 0.2$.  The simulation starts with random initial conditions inside the square whose vertices are the vacua $(\pm \sqrt{15/7},\pm \sqrt{15/7}) $. Note that only stable $Y$-type junctions survive. From left to right and top to bottom, the horizon is approximately 1/16, 1/8, 1/4 and 1/2 of the box size respectively.}
\label{bazeia2positivo}
\end{figure}

\begin{figure}
\begin{center}
\includegraphics*[width=3.6cm, height=3.6cm]{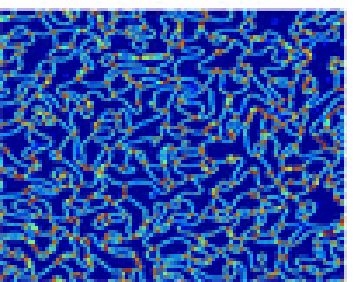}
\includegraphics*[width=3.6cm, height=3.6cm]{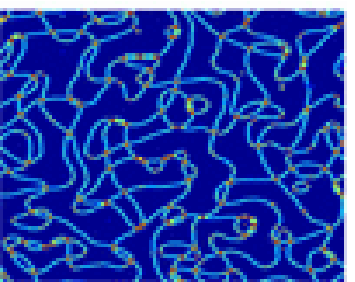}
\includegraphics*[width=3.6cm, height=3.6cm]{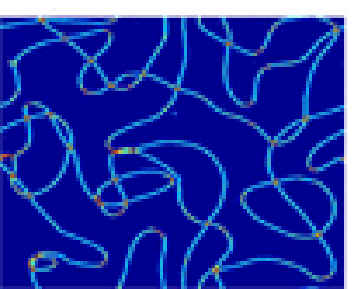}
\includegraphics*[width=3.6cm, height=3.6cm]{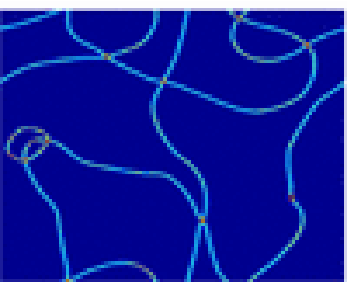}
\end{center}
\caption{Same as Fig. \protect\ref{bazeia2positivo}, except that now $\epsilon = - 0.2$. Here the simulation starts with initial conditions chosen randomly in the square whose vertices are the vacua $(\pm \sqrt{15/13},\pm \sqrt{15/13}) $. Note that in this case only stable $X$-type junctions are present in the network.}
\label{bazeia2negativo1}
\end{figure}

\begin{figure}
\begin{center}
\includegraphics*[width=3.6cm, height=3.6cm]{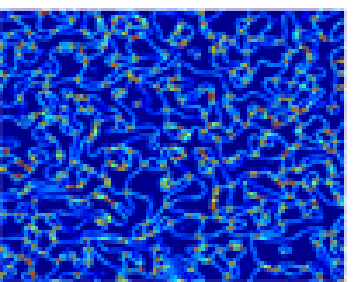}
\includegraphics*[width=3.6cm, height=3.6cm]{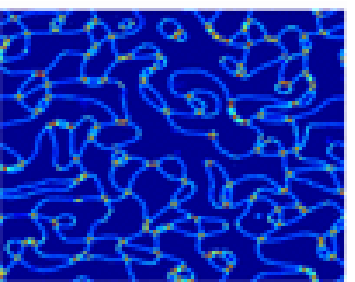}
\includegraphics*[width=3.6cm, height=3.6cm]{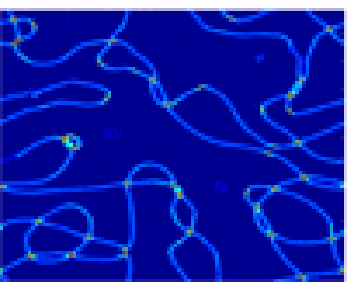}
\includegraphics*[width=3.6cm, height=3.6cm]{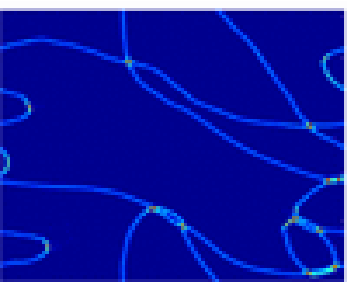}
\end{center}
\caption{Same as Fig. \protect\ref{bazeia2positivo}, except that now $\epsilon = - 0.4$. In this case we start with initial conditions chosen randomly in the square whose vertices are the vacua $(\pm \sqrt{15/7},0) $ and $(0,\pm \sqrt{15/7}) $. Here again only stable $X$-type junctions are present in the network.}
\label{bazeia2negativo2}
\end{figure}

\begin{figure}
\begin{center}
\includegraphics*[width=3.6cm, height=3.6cm]{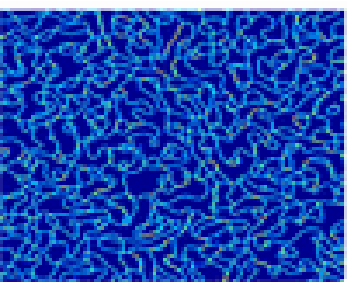}
\includegraphics*[width=3.6cm, height=3.6cm]{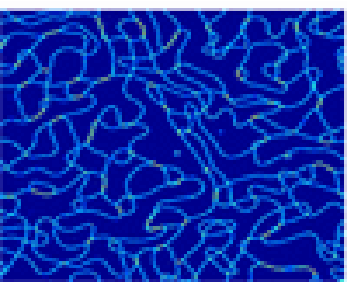}
\includegraphics*[width=3.6cm, height=3.6cm]{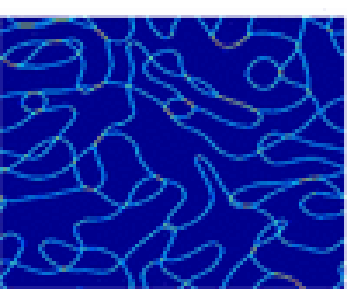}
\includegraphics*[width=3.6cm, height=3.6cm]{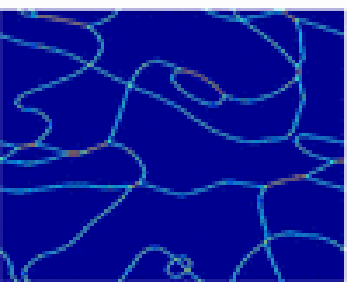}
\end{center}
\caption{Same as Fig. \protect\ref{bazeia2positivo}, except that now $\epsilon = - 0.8$. Here we start with initial conditions chosen randomly in the square whose vertices are the vacua $(\pm \sqrt{15/4},0) $ and $(0,\pm \sqrt{15/4}) $. Note that in this case only $Y$-type junctions are present in the network.}
\label{bazeia2negativo3}
\end{figure}
%%%%%%%%%%%%%%%%%%%%%%%%%%%%%%%%%%%%%%%%%%%%%%%%%%%%%%%%%%%%%%%%%%%%%%

These expectations can be confirmed numerically. This model is very easy to simulate; in what follows we will present the outcome of a few $256^2$ simulations. Although these are relatively small for today's standards, they are more than sufficient to confirm our analysis. For easier comparison with BBL, we will also take $r=\sqrt{3/2}$. In this case, according to the above solution, for  $-1/3 < \epsilon < 2/3$ the vacua are $(\pm (2/3- \epsilon)^{-1/2},\pm (2/3- \epsilon)^{-1/2})$, while for $-4/3 < \epsilon < -1/3$ they are  $(0,\pm (2/3+\epsilon/2)^{-1/2})$ and $(\pm (2/3+\epsilon/2)^{-1/2},0)$. The outcome of two such simulations is illustrated by the four snapshots displayed in Figs. \ref{bazeia2positivo} and \ref{bazeia2negativo1}, where we have respectively taken $\epsilon= \pm 0.2$.\textit{We caution the reader that a pair of Y-type junctions close together might be misidentified as an X-type junction by a careless glance at a low-resolution snapshot.} Notice that indeed in the former case only Y-type junctions survive, and any X-type junction that happens to form is unstable and will decay quickly, while in the latter case it is the X-type junctions that survive.

Again, our treatment is not limited to the asymptotic case $\epsilon \rightarrow 0$, but it is completely general. Our simulations confirm that for any chosen value of $\epsilon > 0$ (that is in the range $0 < \epsilon < 2/3$, given our choice of $r$), the behavior of the system is similar to that in Fig. \ref{bazeia2positivo}: the formation of Y-type junctions is always preferred. On the other hand, we confirm our expectation that two distinct cases appear for $\epsilon <0$. For the interval $-2/3 < \epsilon < 0$, only X-type junctions are formed, as can be seen in Figs. \ref{bazeia2negativo1} and \ref{bazeia2negativo2} for $\epsilon =-0.2$ and $\epsilon =-0.4$ respectively. However, for $\epsilon <-2/3$ X-type junctions are no longer stable, and instead Y-type junctions form, as illustrated in Fig. \ref{bazeia2negativo3} for the case $\epsilon=-0.8$.

%%%%%%%%%%%%%%%%%%%%%%%%%%%%%%%%%%%%%%%%%%%%%%%%%%%%%%%%%
\begin{figure}
\begin{center}
\includegraphics*[width=3.6cm, height=3.6cm]{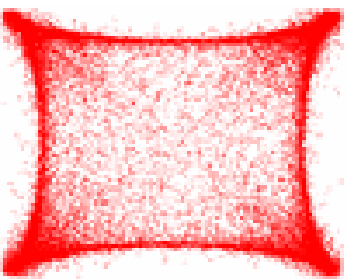}
\includegraphics*[width=3.6cm, height=3.6cm]{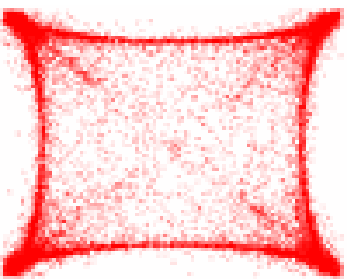}
\includegraphics*[width=3.6cm, height=3.6cm]{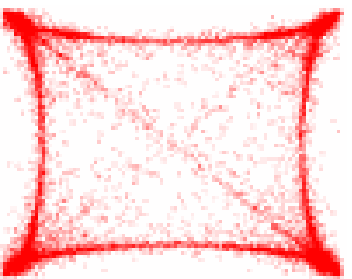}
\includegraphics*[width=3.6cm, height=3.6cm]{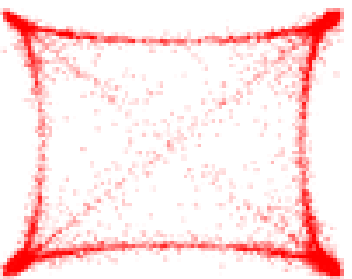}
\end{center}
\caption{The phase space distribution for the four timesteps of the simulation in Fig. \protect\ref{bazeia2positivo}. The vertices of the square in the plane $(\phi_1,\phi_2)$ are the minima of the potential, given by $(\pm \sqrt{15/7}), \pm \sqrt{15/7})$.}
\label{bazeia2positivofase}
\end{figure}

\begin{figure}
\begin{center}
\includegraphics*[width=3.6cm, height=3.6cm]{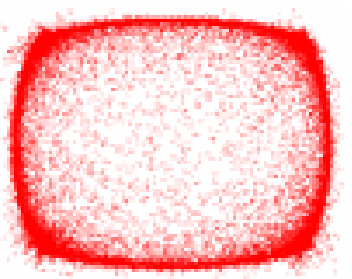}
\includegraphics*[width=3.6cm, height=3.6cm]{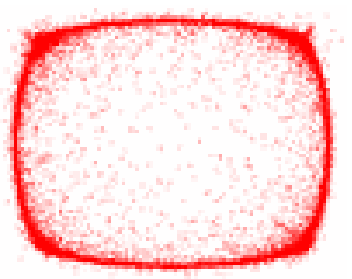}
\includegraphics*[width=3.6cm, height=3.6cm]{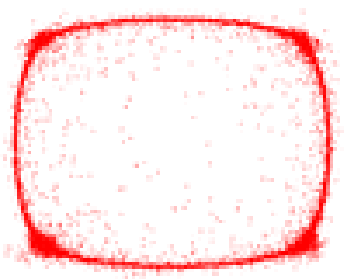}
\includegraphics*[width=3.6cm, height=3.6cm]{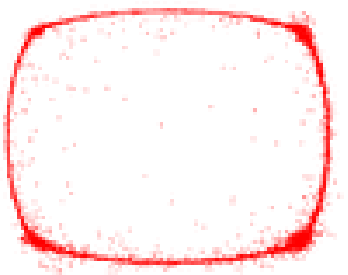}
\end{center}
\caption{The phase space distribution for the four timesteps of the simulation in Fig. \protect\ref{bazeia2negativo1}. The vertices of the square in the plane $(\phi_1,\phi_2)$ are the minima of the potential, given by $(\pm \sqrt{15/13}), \pm \sqrt{15/13})$.}
\label{bazeia2negativofase1}
\end{figure}

\begin{figure}
\begin{center}
\includegraphics*[width=3.6cm, height=3.6cm]{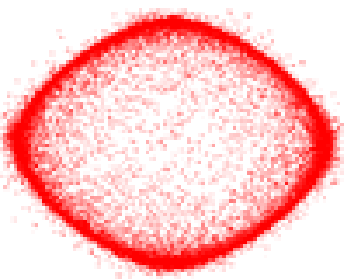}
\includegraphics*[width=3.6cm, height=3.6cm]{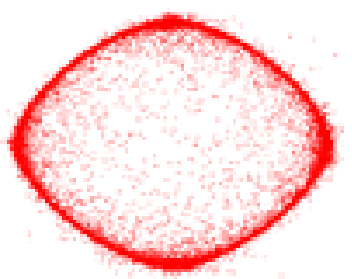}
\includegraphics*[width=3.6cm, height=3.6cm]{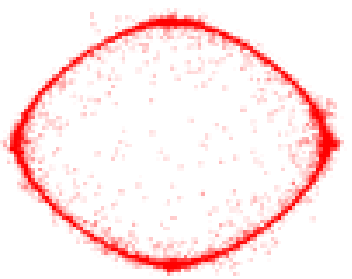}
\includegraphics*[width=3.6cm, height=3.6cm]{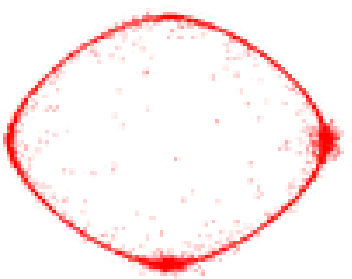}
\end{center}
\caption{The phase space distribution for the four timesteps of the simulation in Fig. \protect\ref{bazeia2negativo2}. The vertices of the square in the plane $(\phi_1,\phi_2)$ are the minima of the potential, given by $(\pm \sqrt{15/7}), 0)$ and $(0, \pm \sqrt{15/7})$.}
\label{bazeia2negativofase2}
\end{figure}

\begin{figure}
\begin{center}
\includegraphics*[width=3.6cm, height=3.6cm]{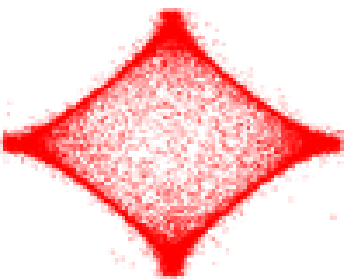}
\includegraphics*[width=3.6cm, height=3.6cm]{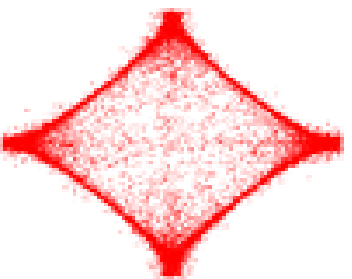}
\includegraphics*[width=3.6cm, height=3.6cm]{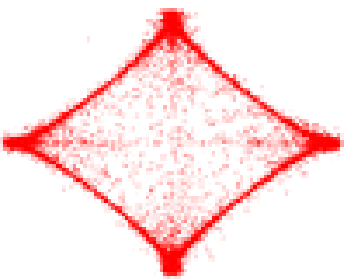}
\includegraphics*[width=3.6cm, height=3.6cm]{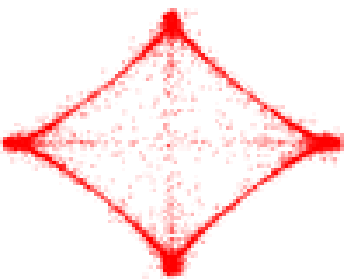}
\end{center}
\caption{The phase space distribution for the four timesteps of the simulation in Fig. \protect\ref{bazeia2negativo3}. The vertices of the square in the plane $(\phi_1,\phi_2)$ are the minima of the potential, given by $(\pm \sqrt{15/4}), 0)$ and $(0, \pm \sqrt{15/4})$.}
\label{bazeia2negativofase3}
\end{figure}
%%%%%%%%%%%%%%%%%%%%%%%%%%%%%%%%%%%%%%%%%%%%%%%%%%%%%%%%%%%%%%%%%%%

As further confirmation, it is also interesting to analyze the behavior of the phase space formed by the two scalar fields, $\phi_1$ and $\phi_2$. Figures \ref{bazeia2positivofase}--\ref{bazeia2negativofase3} show the field configurations corresponding to the snapshots of Figs. \ref{bazeia2positivo}--\ref{bazeia2negativo3} respectively. It is quite clear that in the cases $\epsilon>0$ and $\epsilon <-2/3$ the field values tend to concentrate along the solutions of both the \textit{edges} and the \textit{diagonal} sectors of the square---notice that the orientation of the square is different in the two cases, corresponding to the solutions given by Eq. (\ref{solplus}) and (\ref{solminus}) respectively. On the other hand, in the $-2/3 < \epsilon <0$ range the \textit{diagonal} sector of the square is a repeller (which gradually becomes depopulated) and only the \textit{edges} are populated. These numerical results confirm our discussion above, as well as our original results in Paper I. We emphasize that there is a valuable lesson to be learned at this point: the type of stable junctions that will form (in physical space) in a given model is an issue that is related to, but quite distinct from, the geometry of the various vacua in configuration space. In other words, there are both topological an energetic considerations involved in determining the detailed properties of a given model, and one must be mindful of both of them in any rigorous analysis.

Finally, we should also emphasize that the particular cases where $\epsilon r^2 = 0$ and $\epsilon r^2=-1$ do not represent marginal cases between the formation of the junctions of the type $Y$ and $X$. Instead, these choices decouple the two fields, and so no junctions are formed anymore. Consequently simulations will simply show the superposition of two single fields evolving independently. This is trivial to see for the case $\epsilon r^2 = 0$. For the case $\epsilon r^2=-1$, this is best seen by defining a new pair of scalar fields $\varphi_1$ and $\varphi_2$ as
\bq
\varphi_1=\frac1{\sqrt{2}}\,\left( \phi_1 + \phi_2 \right)\,\\
\varphi_2=\frac1{\sqrt{2}}\,\left( \phi_1 - \phi_2 \right)\,
\eq
and writing the potential in terms of these. One then finds that these rotated fields decouple for $\epsilon r^2=-1$. 

For an interesting example of a marginal case where both types of junctions are allowed, we can consider the model described by a complex scalar field $\Phi$ with Lagrangian
\be
\mathcal{L} = \partial_\mu \Phi \partial^\mu \bar{\Phi} - V(\Phi)
\ee
and the potential
\be
V(\Phi)=\kappa \left|\Phi^N-1\right|^2\,;\label{pot2}
\ee
where $\kappa$ is a real parameter and $N$ is integer. This model has vacua at
$$
\Phi=e^{i\frac{n}{N}},\quad n=0,1,\ldots,N-1\,.
$$
We can alternatively define the field's phase as $\phi$, and it is then obvious that we will have $N$ minima, evenly spaced around $\phi$. The case $N=2$ produces standard domain walls and $N=3$ produces Y-type junctions, but the case $N=4$ is slightly more subtle. The potential (\ref{pot2}) has supersymmetric properties, and hence the energy of a specific solution depends only on the initial and final vacua. In other words, the possible ways of connecting two opposite vacua (directly or through the intermediate vacuum) will have the same energy. Hence this case is an example of the scenario where $\sigma_d = 2\,\sigma_e$. There is therefore no local energetic argument preferring one type of junction to the other, and consequently Y-type and X-type junctions will always co-exist. This can also be confirmed numerically, and visually the results can be approximately described as the superposition of the two distinct cases shown above.

%%%%%%%%%%%%%%%%%%%%%%%%%%%%%%%%%%%%%%%%%%%%%%%%%%%%%%%%%%%%%%
\section{\label{threefield}Three-Field Models}

We now move on to the case of models with three scalar fields. The discussion turns out to be quite similar to the two-field case, though at some points the effect of the increased dimensionality provides for different phenomenology. We will again consider the analogous BBL model \cite{BAZEIA}, but we will also discuss the perturbed $O(3)$ Kubotani model, which has been studied in the past, starting with \cite{KUBOTANI}. In fact we shall show that the two models are very similar, though not quite identical.

\subsection{The BBL model}

The extension of the BBL model of \cite{BAZEIA} for three real scalar fields is straightforwardly given by
\be
\mathcal{L} = \frac12 \sum_{i=1}^3 (\partial_\mu \phi_i \partial^\mu \phi_i)  + V(\phi_i),
\ee 
where the $\phi_i$ are real scalar fields and the potential has the form
\bq
V(\phi_i) &=& \frac12 \sum_{i=1}^3 \left[(r-\frac{\phi_i^2}{r})^2 + \epsilon (\phi_i^4 + \frac92) \right]\nonumber \\
 &-& 3 \epsilon\left( \phi_1^2 \phi_2^2 + \phi_1^2 \phi_3^2 +\phi_2^2 \phi_3^2 \right)\,;
 \label{bbl3}  
\eq
where again $r$ and $\epsilon$ are two real parameters. As in the two-field case, there are two branches for the minima. In this case for $ -2/5 < \epsilon r^2 <  1/2$  the minima are of the form
\be
\phi_i^2=\frac{r^2}{1 - 2 \epsilon r^2},\quad i=1,2\,,\label{solplus3}
\ee
while for $ - 1 < \epsilon r^2 < - 2/5$ they are of the form
\be
\phi_i^2 = \frac{r^2}{1+\epsilon r^2}, \quad \phi_{j \neq i}^2=0\,.\label{solminus3}
\ee
In the former case there are 8 minima, which are placed at the vertices of a cube in the space $(\phi_1,\phi_2,\phi_3)$. In the latter one, there are 6 minima, which are placed at the vertices of an octahedron. The minima of the second case can alternatively be thought of as being located at the centers of the faces of a cube.

In the first case (\ref{solplus3}), there are twenty-eight topological sectors and three kinds of walls, which for obvious reasons we can refer to as \textit{edges}, \textit{external diagonals} and \textit{internal diagonals}. The number of different walls of each type is respectively twelve, twelve and four. In the second case (\ref{solminus3}), there are fifteen topological sectors and two kinds of walls, which we can refer to as \textit{edges} and \textit{axes}. The number of different walls of each type is respectively twelve and three. 

\begin{figure}
\begin{center}
\includegraphics*[width=6.9cm]{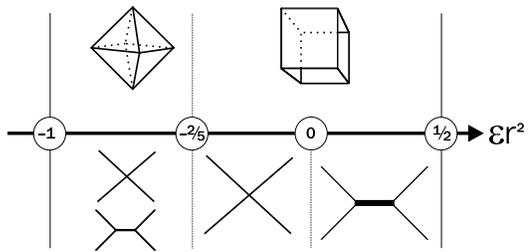}
\end{center}
\caption{Types of junctions (bottom) and configuration of the minima (top), as a function of the parameter $\epsilon r^2$ in the two-field BBL model \protect\cite{BAZEIA}. Notice that in the case $\epsilon r^2>0$ there are effectively two kinds of Y-type junctions, involving either the \protect\textit{internal diagonal} or the \protect\textit{external diagonal} sector.}
\label{config3d}
\end{figure}

Again the choice of the parameter $\epsilon$ determines what type of junctions will be present. For the branch $\epsilon r^2 >-2/5$, corresponding to the cubic solution of Eq. (\ref{solplus3}), X-type junctions survive only if the junctions involving walls from the diagonal sectors are energetically disfavored. This requires $\sigma_{id} > 3 \sigma_{e}$ (for the internal diagonals) and $\sigma_{ed} > 2 \sigma_{e}$ (for the external ones). These conditions are only verified in the interval $-2/5 < \epsilon r^2< 0$, so in this range we do have stable X-type junctions. Conversely, for $0 < \epsilon r^2 < 1/2$ only the Y-type junctions survive, and these may involve either of the \textit{diagonal} sectors, so we effectively have two types of Y-junctions. On the other hand, for the branch $\epsilon r^2 <-2/5$, corresponding to the octahedral solution of Eq. (\ref{solminus3}), both Y-type and X-type junctions can survive. Note that in this octahedron branch X-type junctions, as well as any Y-type junctions which involve the \textit{axes} sector (as opposed to only the \textit{edge} sector), correspond to field space configurations where one of the three fields vanishes. The case $\epsilon = 0$ is again not interesting because it represents the evolution of the three decoupled fields, with no junctions being formed. A numerical study can be performed along the same lines as what was done for the two-field case, and its outcome is summarized in Fig. \ref{config3d}.

%%%%%%%%%%%%%%%%%%%%%%%%%%%%%%%%%%%%%%%%%%%%%%%%%%%%%%%%%
%%%%%%%%%%%%%%%%%%%%%%%%%%%%%%%%%%%%%%%%%%%%%%%%%%%%%%%
\subsection{The Kubotani model}

The perturbed $O(N)$ models are a different class of models with N real scalar fields, which have been studied in \cite{KUBOTANI,BATTYE}. Since the case that has been studied in more detail is $N=3$, it is interesting to contrast it with the BBL model. In full generality, the model has the following potential
\be
V(\phi)=\lambda\left(\sum_{i=1}^N \phi_i^2-\eta^2\right)^2 + \xi \sum_{i=1}^N\left(\phi_i^2-\zeta^2\right)^2;\label{onpot}
\ee
the model parameters should be such that $\xi + \lambda>0$ and $\xi+ N\lambda>0$.

A few particular cases are worth pointing out. Firstly, the case $\lambda=0, \xi>0$ is the decoupled limit. This has minima $\phi_i^2=\zeta^2$. Again there are no real junctions: the different types of walls simply pass through each other without interacting, whereas walls of the same type intercommute normally. Still this is interesting as a test for numerical codes.
Secondly, the case $\xi=0, \lambda>0$ is the $O(N)$-symmetric case. Here the minima are $\phi_i^2=\eta^2/N$. The N-th case has a total of $2^N$ minima, corresponding to the vertices of an N-dimensional hypercube.

\begin{figure}
\begin{center}
\includegraphics*[width=3.6cm,height=3.6cm]{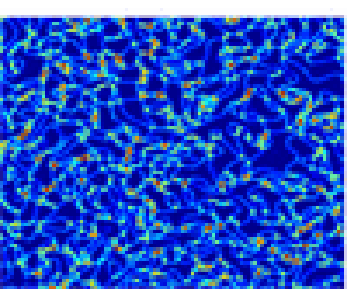}
\includegraphics*[width=3.6cm,height=3.6cm]{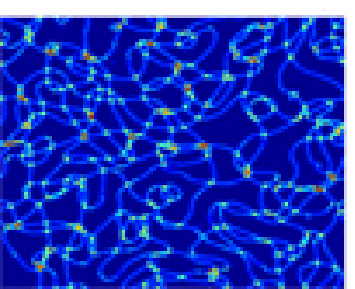}
\includegraphics*[width=3.6cm,height=3.6cm]{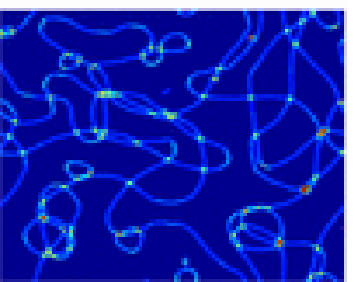}
\includegraphics*[width=3.6cm,height=3.6cm]{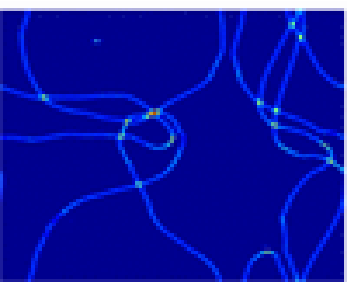}
\end{center}
\caption{The matter-era evolution of the domain wall network of a Kubotani-type model with a parameter choice $\lambda=3/20$, $\eta^2=10/3$ and $\xi=1/12$. This also mimics a three-field BBL model with $r=\sqrt{3/2}$ and $\epsilon = -0.2$.  Note that only stable X-type junctions survive. From left to right and top to bottom, the horizon is approximately 1/16, 1/8, 1/4 and 1/2 of the box size respectively.}
\label{kuboplus}
\end{figure}

In the general case the minima are of the form
\be
\phi^2_i=\frac{\lambda \eta^2+\xi \zeta^2}{N\lambda+\xi},\quad i=1,\ldots, N\,,\label{solkuboplus}
\ee
for the case $\xi\ge0$ (as given in \cite{BATTYE}), or
\be
\phi^2_i=\frac{\lambda \eta^2+\xi \zeta^2}{\lambda+\xi},\quad \phi^2_{j\neq i}=0\,,\label{solkubominus}
\ee
for the case $\xi <0$. The analogies with the BBL model should be fairly clear. In general the former has $2^N$ minima, whose vertices form an N-dimensional (hyper)cube, while the latter has $2N$ minima, that can be thought of as lying in the center of the faces of the said N-dimensional cube. The wall thickness is
\be
\delta\sim |\xi|^{-1/2}\eta^{-1},
\ee
while the wall tension is
\be
T\sim |\xi|^{1/2}\eta^3\,.
\ee
Another particular subclass of models has $\zeta=0$ and $N=3$; these were first studied in \cite{KUBOTANI}, and for this reason are usually referred to as Kubotani models, and we will use this choice for our numerical illustrations. The key difference between this class of models and the BBL models is the fact that for $\xi>0$ the tension of the \textit{diagonal} walls (of which there are $(N-1)$ types in the $N$-dimensional case) is always higher than $N$ times the value of the \textit{edge} walls. Consequently, for $\xi>0$ the stable junctions will always be of X-type. In other words, unlike the BBL model, Y-type junctions will never be stable in the cubic branch. The results of one simulation in this regime are shown in Fig. \ref{kuboplus}, where we have chosen $\lambda=3/20$, $\eta^2=10/3$ and $\xi=1/12$.

%%%%%%%%%%%%%%%%%%%%%%%%%%%%%%%%%%%%%%%%%%%%%%%%%%%%%%%%%

\begin{figure}
\begin{center}
\includegraphics*[width=3.6cm,height=3.6cm]{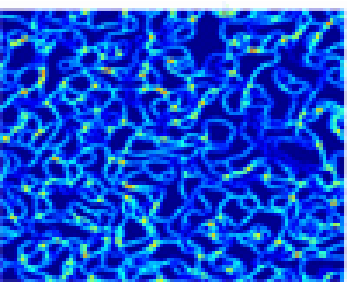}
\includegraphics*[width=3.6cm,height=3.6cm]{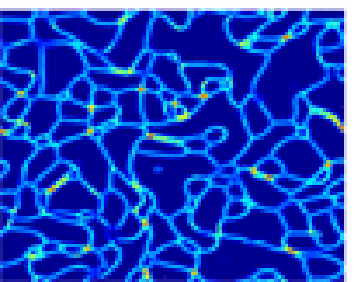}
\includegraphics*[width=3.6cm,height=3.6cm]{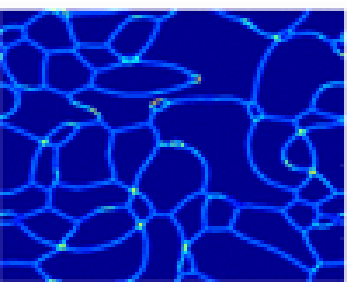}
\includegraphics*[width=3.6cm,height=3.6cm]{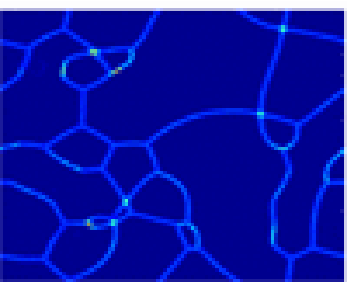}
\end{center}
\caption{
The matter-era evolution of the domain wall network of a Kubotani-type model with a parameter choice $\lambda=3/10$, $\eta^2=5/3$ and $\xi=-1/6$. This also mimics a three-field BBL model with $r=\sqrt{3/2}$ and $\epsilon = -0.4$.  Note that both Y-type and X-type junctions survive. From left to right and top to bottom, the horizon is approximately 1/16, 1/8, 1/4 and 1/2 of the box size respectively.}
\label{kubominus}
\end{figure}
%%%%%%%%%%%%%%%%%%%%%%%%%%%%%%%%%%%%%%%%%%%%%%%%%%%%%%%%%%%%%

For $\xi < 0$, it is again clear that Y-type junctions will form. But just like in the BBL case, one can also envisage having X-type junctions, which can arise from field space configurations where one of the three scalar fields vanishes. Notice that in fact a chessboard-type lattice could be built, with just four of the vacua lying on a plane, that might be stable. However, no realistic spontaneous symmetric breaking mechanism can occur that selects only a subset of the vacua. Therefore the existence of both kinds of junctions is expected in any realistic simulation. This fact can be confirmed in Fig. \ref{kubominus} for the parameter choices $\lambda=3/10$, $\eta^2=5/3$ and $\xi=-1/6$. We also note that even though the majority of the junctions are of Y-type, the fraction of X-type junctions remains approximately constant during the evolution. This issue deserves further study.

%%%%%%%%%%%%%%%%%%%%%%%%%%%%%%%%%%%%%%%%%%%%%%%%%%%%%%%%%
 \begin{figure}
\begin{center}
\includegraphics*[width=3.6cm,height=3.6cm]{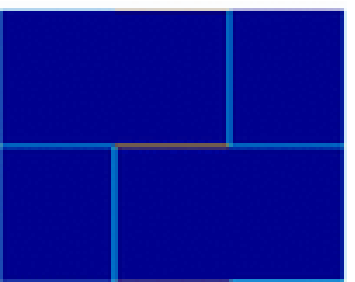}
\includegraphics*[width=3.6cm,height=3.6cm]{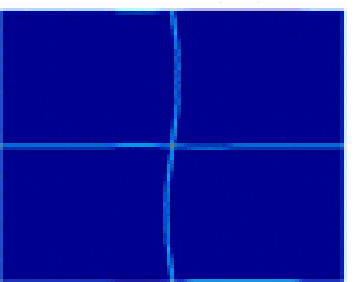}
\end{center}
\caption{Two Y-type junctions in a Kubotani-type $\xi<0$ model decay into a stable X-type junction; see the text for a detailed discussion.}
\label{kubo3to4}
\end{figure}

\begin{figure}
\begin{center}
\includegraphics*[width=3.6cm,height=3.6cm]{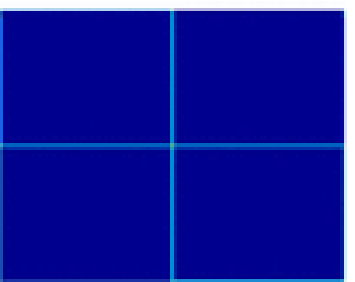}
\includegraphics*[width=3.6cm,height=3.6cm]{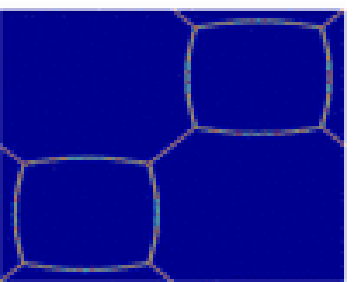}
\end{center}
\caption{An X-type junction in a Kubotani-type $\xi<0$ model decays into two Y-type junctions; see the text for a detailed discussion.}
\label{kubo4to3}
\end{figure}
%%%%%%%%%%%%%%%%%%%%%%%%%%%%%%%%%%%%%%%%%%%%%%%%%%%%%%%

This can be seen more easily if we construct by hand a simple square network containing four distinct vacua. In Fig. \ref{kubo3to4} we start with four vacua lying on a plane and forming two Y-type junctions: these are destroyed and a new X-type junction is formed. Conversely, in Fig. \ref{kubo4to3} we have a set of four vacua that are not on a plane in field space: in this case any X-type junction is unstable, and decays into a pair of Y-type ones.

\subsection{Relating the BBL and Kubotani models}

The above discussion makes it clear that the BBL and Kubotani models are quite similar, but it is actually possible to exhibit a direct relation between both potentials. The easiest way to see this is to relate them through the generalized potential
\be
V = A \sum_i \phi_i^4 + B \sum_i \phi_i^2 + C \sum_{i \neq j} \phi_i^2 \phi_j^2 + D.\label{genericpot}
\ee
where $A$, $B$, $C$ and $D$ are real parameters. For the values for the parameters given by
\bq
A\,&=&\,\frac{1}{2 r^2}+\frac{\epsilon}{2} \quad \,B\,=\,-1 \\ 
C\,&=&\,-\frac{3 \epsilon}{2} \,\,\,\,\,\quad D\,=\,\frac{r^2}{2}+\frac{27 \epsilon}{4}\label{coefbbl}
\eq
one has exactly the potential of the BBL model. On the other hand, the choice
\bq
A\,&=&\,\lambda+\xi,\,\,\,\quad B\,=\,-2 \lambda \eta^2,\\
C\,&=&\,2 \lambda,\,\,\,\quad D\,=\,\lambda \eta^4\label{coefkubo}
\eq
leads to the potential for the Kubotani model (recall that this has $\zeta=0$).

Using the relations above, it is easy to see that the models are related by
\bq
\lambda = -\frac{3 \epsilon}{4},\,\,\,\quad \eta^2= -\frac{2}{3\epsilon}, \,\,\,\quad \xi= \frac{1}{2 r^2} + \frac{5 \epsilon}{4},\label{coefrel}
\eq
where $\lambda$, $\xi$ and $\eta$ are the Kubotani parameters and $\epsilon$, $r$, are the corresponding BBL parameters. The relations above lead us to conclude that for $\lambda >0$ and $\eta^2 >0$, only the interval $\epsilon < 0$ of the BBL model can be mapped into the Kubotani model. In addition, we see that $\xi >0$ is related to the range $-2/5 < \epsilon r^2 < 0$ while $\xi<0$ is related to $-1/2 < \epsilon r^2 < -2/5$. However, notice that there are two consistency conditions, $2\eta\lambda^2=1$ on the Kubotani side, and $\epsilon r^2=-27\epsilon^2/2-2/3$ on the BBL side (though the latter affects only the constant term in the potential), so even in this restricted $\epsilon <0$ range the correspondence is not one-to-one.

\begin{figure}
\begin{center}
\includegraphics*[width=3.6cm,height=3.6cm]{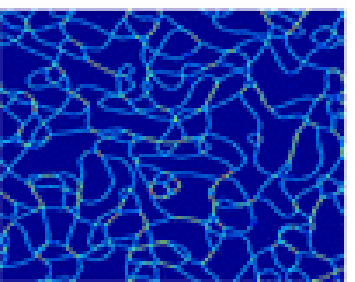}
\includegraphics*[width=3.6cm,height=3.6cm]{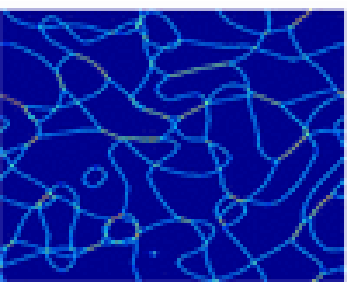}
\includegraphics*[width=3.6cm,height=3.6cm]{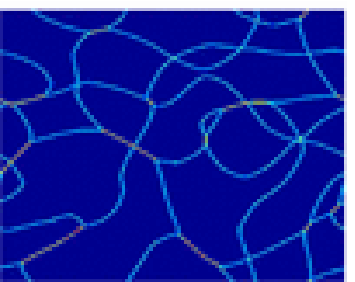}
\includegraphics*[width=3.6cm,height=3.6cm]{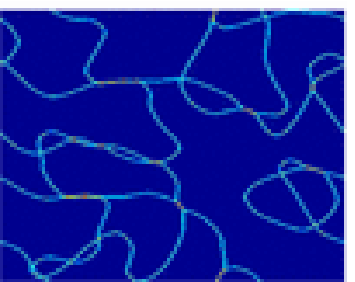}
\end{center}
\caption{
The matter-era evolution of the domain wall network for a three-field BBL model with $r=\sqrt{3/2}$ and $\epsilon = 0.1$.  Note that only stable Y-type junctions survive. From left to right and top to bottom, the horizon is approximately 1/16, 1/8, 1/4 and 1/2 of the box size respectively.
}
\label{bazeia3cubeY}
\end{figure}

Hence our Kubotani-model simulations in Figs. \ref{kuboplus} and \ref{kubominus} can also be interpreted as BBL model simulations. In Fig. \ref{kuboplus} we took $\lambda=3/20$, $\eta^2=10/3$ and $\xi=1/12$. Note that only X-type junctions are formed. Using the relations above we find that this corresponds to the BBL model for $r=\sqrt{3/2}$ and $\epsilon=-0.2$, and similar results are obtained for the entire interval $-2/5 < \epsilon r^2< 0$. On the other hand, in Fig. \ref{kubominus} we have used $\lambda=3/10$, $\eta^2=5/3$ and $\xi=-1/6$ in the Kubotani model ($r=\sqrt{3/2}$ and $\epsilon=-0.4$ in BBL model). Only Y-type junctions survive for any chosen negative value of $\xi$ (that is $-1/2 < \epsilon r^2< -2/5$). Finally, for completeness we present the case for $0 < \epsilon r^2< 1/2$ in the BBL model, which has no correspondence in the Kubotani case. We have chosen $r=\sqrt{3/2}$ and $\epsilon=0.1$ to illustrate this scenario in Fig. \ref{bazeia3cubeY}. 

%%%%%%%%%%%%%%%%%%%%%%%%%%%%%%%%%%%%%%%%%%%%%%%%%%%%%%%%%
%%%%%%%%%%%%%%%%%%%%%%%%%%%%%%%%%%%%%%%%%%%%%%%%%%%%%%%%%%

\section{The Ideal Model}

The availability of a large number of models (of which the two studied above are but examples), all different but to some extent related, begs the question of which features are fundamental and which are irrelevant details. More to the point, one might ask what is the best possible model for domain walls, at least from the point of view of its potential to produce frustrated networks. Based on our results so far, we shall discuss what the key characteristics of such a model are, and discuss an explicit construction. It turns out that a model considered in \cite{CARTER} has some of the characteristics of this ideal model. In order to make the argument clear, we shall go through the key steps in considerable detail, starting with a discussion we already presented in Paper I and building up from there.

In a two-dimensional domain wall network generated dynamically by a cosmological phase transition, the number of edges of the domains can assume several values, with a particular distribution. The dimensionality of the junctions in such a network, that is, the number of walls meeting each junction, depends on the structure of the chosen potential. One example is the Kubotani model that can be seen in Figs. \ref{kuboplus}(for $\xi >0$) and \ref{kubominus} (for $\xi <0$). While in the former case only X-junctions are present, the latter one presents both Y-type and X-type junctions. Note in both cases the stable junctions are formed by walls with equal tension. We should emphasize that as the average dimensionality of the junctions increases, the average number of faces of the domains will be smaller.

%%%%%%%%%%%%%%%%%%%%%%%%%%%%%%%%%%%%%%%%%%%%%%%%%%%%%%%%%%%%%
\begin{figure}
\begin{center}
\includegraphics*[width=5.5cm]{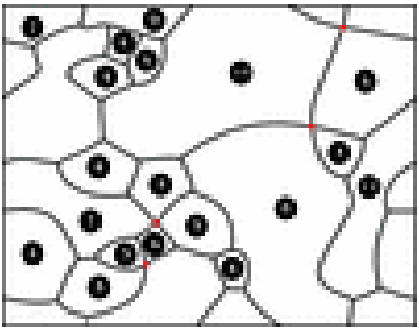}
\end{center}
\caption{Illustration of a random domain distribution in a planar network taken from the fourth snapshot of Fig. \ref{kubominus}. Each domain has $x$ edges and each junction is formed by $d$ walls. The average values are $\langle d \rangle = 53/17 $ and $\langle x \rangle = 106/19$.
}
\label{euler}
\end{figure}
%%%%%%%%%%%%%%%%%%%%%%%%%%%%%%%%%%%%%%%%%%%%%%%%%

As a simple illustration, we can take the fourth snapshot in Fig. \ref{kubominus}, which is schematically shown in Fig. \ref{euler}. The average number of faces $x$ of each domain is related to the dimensionality $d$ of the junctions by 
\be
\left( \langle x \rangle -2 \right)\left( \langle d \rangle- 2 \right)\,=\,4.
\label{eulerequation}
\ee
In Fig. \ref{euler}, these values are $\langle d \rangle = 53/17\sim3.11 $ and $\langle x \rangle = 106/19\sim5.58$.
We have also verified that these numbers are approximately constant during the evolution of the network. Indeed, one can in principle use these numbers to characterize each particular model, that is each choice of the potential and of the corresponding free parameters. This is another issue that is beyond the scope of the present discussion, but to which we shall return in the future.

Now, the stability of each particular domain will depend, among other things, on its number of edges. Let us momentarily assume that $d$ is fixed to be equal to $3$. By using the Euler relation in Eq. \ref{eulerequation}, such network must have $\langle x \rangle = 6$. Of course the network will contain domains with different values of $x$. However we have shown in Paper I  that, in this case, only domains with $x \ge 6$ might possibly be stable, and that all domains with $x<6$ will quickly collapse in order to minimize the energy. Therefore, no domain wall evolution starting by random initial conditions in such models can provide a stable network. Furthermore, in scenarios where there are domain walls with different tensions, an additional source of instability is also present: when these heavier walls collapse, the dimensionality of the corresponding junction is increased. 

Putting all these facts together, the logical consequences are clear. In an ideal model, the probability that two domains in the same vacuum state are close to each other should be as small as possible. This can be accomplished in models with a very large number of different vacua (say, having the number of scalar fields involved $N \to \infty$). In this case the collapse of a single domain will only very rarely lead to the fusion of two of the surrounding domains. Note that this fusion would lead to a further reduction of the total number of edges of the contiguous domains which as we saw is clearly undesirable since it would increase the probability that some of the contiguous domains would themselves become unstable to collapse. On the other hand, all the possible domain walls should have equal tensions. Again, if that were not the case we would be adding a different source of instability since the walls with higher tension would tend to collapse, thereby increasing the dimensionality of the junctions which, in turn, would lead to a decrease in the average number of edges of individual domains and to the production of further unstable two-edge domains.

The ideal model (as far as frustration is concerned) is therefore a model with a very large number of vacua with all the domain walls connecting the various vacua having the same tension. In Paper I we have already shown that, due to energetic considerations, the stable junctions of such a model must necessarily be of Y-type only, so that $d=3$. We have also shown that the only possible equilibrium configuration in two dimensions with Y-type junctions is the 'honeycomb' hexagonal-type lattice (otherwise unstable two, three, four and five edge domains would necessarily occur). The fact such lattices are not really stable (they allow for locally confined energy deformations although the Hubble damping may temporarily prevent the collapse of such configurations) together with the very small probably that such a lattice could be an attractor
for the evolution of domain wall network simulations from realistic initial conditions has led us to our no frustration conjecture. However, the analysis of Paper I can only be fully applied to domain wall networks in two spatial dimensions, and therefore the construction of an explicit realization of the ideal model is crucial if we want to test our conjecture in the more realistic case of three spacetime dimensions. It is this missing step that we now provide.

Geometrically speaking, the ideal potential is therefore described by $N$ real scalar fields and has mutually equidistant minima. The number of vacua is $N+1$ for $N$ real scalar fields, and the energetic cost for a specific transition between any two of them is the same. An explicit realization of such a model is given by the potential
\be
V=\frac{\lambda}{N+1}\sum_{j=1}^{N+1} r_j^2 \left(r_j^2 - r_0^2\right)^2 \label{ideal}
\ee
with
\be
r_j^2=\sum_{i=1}^N (\phi_i - p_{{i}_{j}})^2
\ee
where $p_{{i}_{j}}$ are the $N+1$ coordinates of the vacua of the potential. We have chosen $p_{{i}_{j}}$ to be the vertices of an ($N+1$)-dimensional regular polygon, and fixed the distance between the vacua to be equal to the parameter $r_0$. The model is therefore the sum of $N+1$ $\phi^6$ potentials. Each of them has one minimum located at the center, and a continuum set of minima at a distance $r_0$ from the center. Note that $N$ of these vacua are located exactly at the centers of the other potentials.

%%%%%%%%%%%%%%%%%%%%%%%%%%%%%%%%%%%%%%%%%%%%%%%%%%%
\begin{figure}
\begin{center}
\includegraphics*[width=4.2cm]{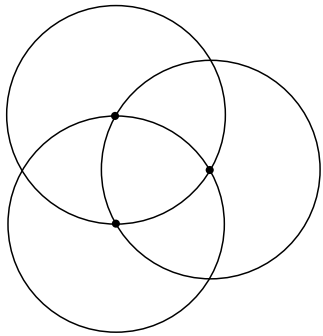}
\end{center}
\caption{The illustration of the vacua displacement of three Higgs potential. Note that there are only three equidistant points where these minima coincide.}
\label{three}
\end{figure}

\begin{figure}
\begin{center}
\includegraphics*[width=4.2cm]{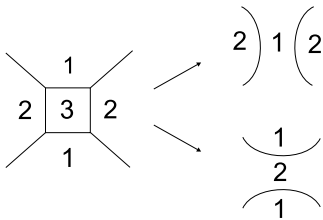}
\end{center}
\caption{
Possible collapses of a domain with four edges in the model with only three vacua.}
\label{colapso1}
\end{figure}

\begin{figure}
\begin{center}
\includegraphics*[width=4.2cm]{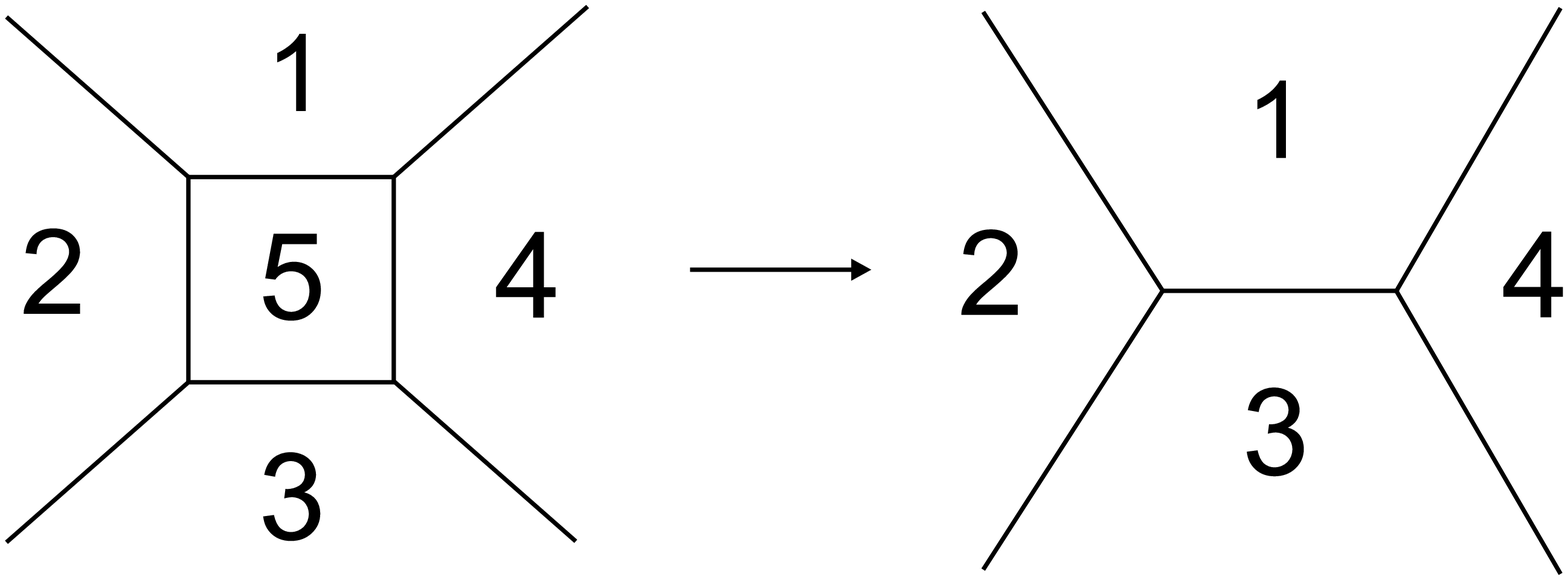}
\end{center}
\caption{
Possible collapse of a domain with four edges in the model with more than five vacua.}
\label{colapso2}
\end{figure}
%%%%%%%%%%%%%%%%%%%%%%%%%%%%%%%%%%%%%%%%%%%%%%%%%%%%%%%%%

For instance, if one takes $N=2$ here we will have three equidistant vacua at the vertices of an equilateral triangle, as illustrated in Fig. \ref{three}. This particular case is in fact analogous to the model given by Eq. (\ref{pot2}) if one takes $N=3$ in that potential. Despite the guarantee that the only stable junction will be of Y-type, we do not expect that a stable lattice will emerge from random initial conditions. Since this model has only three vacua, the collapse of domains with four edges will always happen as illustrated in Fig. \ref{colapso1}. In some sense, this $N=2$ realization of Eq. (\ref{ideal}) is maximally unstable. It is obvious that if this potential had two more minima, the decay shown in Fig. \ref{colapso2} would become possible. In this configuration the collapse of the square occurs decreasing the number of edges of the neighbor domains, but not cutting the network. Notice that the probability of appearance of such a configuration will still be very small, though enhanceable by the addition of further vacua. Again, this is the reason why we want to have $N\to\infty$.

The case with $N=3$ has four vacua located at the vertices of a tetrahedron. All the minima are connected to each other by walls with equal tension. We emphasize that although the two-field BBL model also has four minima, the two models are completely different. In the BBL case stable X-type junctions may be formed depending on the parameter $\epsilon$, and when there are Y-type junctions two different kinds of wall exist, with different tensions. Here only Y-type junctions are formed and all of them have the same tension. 

%%%%%%%%%%%%%%%%%%%%%%%%%%%%%%%%%%%%%%%%%%%%%%%%%%%%%
 \begin{figure}
\begin{center}
\includegraphics*[width=4.2cm]{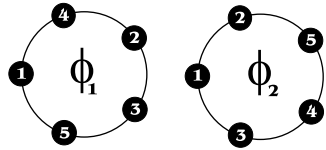}
\end{center}
\caption{
The vacua of the Carter model \protect\cite{CARTER}, in terms of the two phase angles $\phi_1$ and $\phi_2$.}
\label{phases}
\end{figure}
%%%%%%%%%%%%%%%%%%%%%%%%%%%%%%%%%%%%%%%%%%%%%%%%%%%%%

Let us point out a final example, $N=4$, which has four fields whose five minima are located at the vertices of a pentahedron. A very similar analogue was introduced by Carter in Ref. \cite{CARTER}. This indeed has five evenly spaced minima, and all the walls are of the same energy---the only difference is that Carter's model is of $\phi^4$ type, while ours is of $\phi^6$ type. Obviously, both this model and ours allow for the configuration in Fig. \ref{colapso2}. However, as we have discussed, this configuration is not enough for ensuring the stability of network frustration. Note that we again see that Carter's claim that this model will form stable X-type junctions is manifestly incorrect---we will also confirm this numerically below. 

Note that it is also possible to provide a generalization of the Carter model, for the case of a generic (though even) number of real scalar fields. In \cite{CARTER} it is pointed out that the model can be thought of in terms of either four real scalar fields or two complex scalar fields. Indeed, the phases (let's call them $\psi_1$ and $\psi_2$) of these two complex scalar fields provide the simplest way of visualizing the configuration of the vacua---the alternative of $(N+1)$ evenly spaced points on and $(N-1)$-sphere being obviously not very clear. Since the minima must be evenly spaced, they will lie on the surface of some sphere, and one can therefore move between them by moving along some suitably chosen phases. For $N=2M$ real scalar fields ($M$ complex fields) there will therefore be $M$ such phases. Let us consider each such phase individually: it will have a certain number of vacua, equally spaced along it, but obviously only two nearest neighbors. But since there are $M$ such phases, we immediately see that the total number of equidistant minima should be $2M+1=N+1$. In Fig. \ref{phases} we can see the configuration of the minima for the Carter model \cite{CARTER}. Notice that starting at any of the minima, we can reach any of the other minima in a single direct displacement, either moving along $\psi_1$ or along $\psi_2$ as the case may be.

A generalization of the potential given in \cite{CARTER}, which is always of $\phi^4$ type, is the perturbed $[U(1)]^M$ model which has the following form
\be
V(\Phi_i)=\sum_{i=1}^M\left(\Phi_i^2-\eta^2\right)^2+\epsilon^2f_M(\phi_i)\left(\prod_{i=1}^M\Phi_i^2\right)^{2/M}\,.\label{newcarter}
\ee
Here $\Phi_i$ are the $M$ complex scalar fields, with $\phi_i$ the corresponding phases, and $f_M(\phi_i)$ is a function of these phases, having $2M+1$ minima and taking the value $f_{Mmin}=-M$ at those points. The minima of the potential will then be of the form \be
\Phi^2=\frac{\eta^2}{1-\epsilon}\label{cartersol}
\ee
and be evenly spaced for a suitable choice of the function $f_M$. Specifically, for $M=1$ we trivially have
\be
f_1=\cos{(3\phi_1)}\,,\label{carterm1}
\ee
yielding the three evenly spaced minima we already mentioned, while for $M=2$ we should choose
\be
f_2=\cos{(2\phi_1+\phi_2)}+\cos{(2\phi_2-\phi_1)}\,,\label{carterm2}
\ee
which is the potential discussed in \cite{CARTER}, and yields the five evenly spaced minima shown in Fig. \ref{phases}. The construction of generalizations for higher values of $M$ is straightforward in principle, though algebraically tedious in practice.

%%%%%%%%%%%%%%%%%%%%%%%%%%%%%%%%%%%%%%%%%%%%%%%%%%%%%%%%
\begin{figure}
\begin{center}
\includegraphics*[width=3.6cm,height=3.6cm]{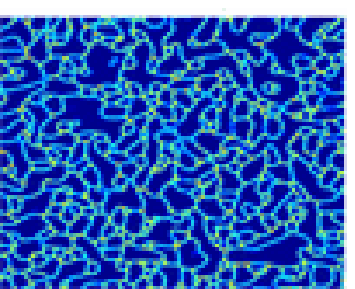}
\includegraphics*[width=3.6cm,height=3.6cm]{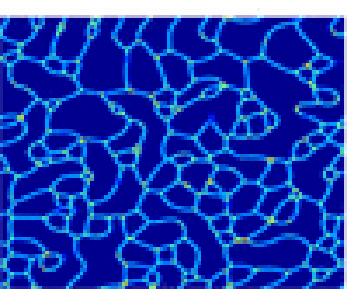}
\includegraphics*[width=3.6cm,height=3.6cm]{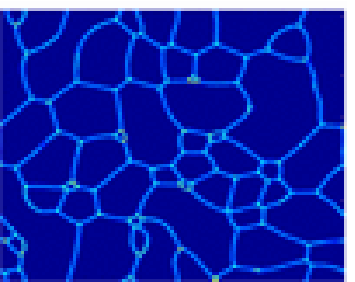}
\includegraphics*[width=3.6cm,height=3.6cm]{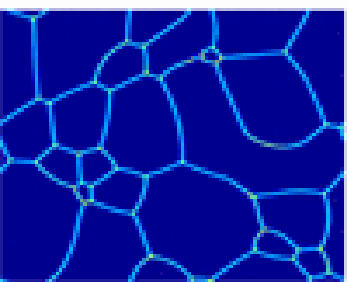}
\end{center}
\caption{
The matter-era evolution of the domain wall network for the ideal model with $N=4$. Note that this model is effectively analogous to that of \protect\cite{CARTER}, and that only stable Y-type junctions survive. From left to right and top to bottom, the horizon is approximately 1/16, 1/8, 1/4 and 1/2 of the box size respectively.
}
\label{ideal4}
\end{figure}

\begin{figure}
\begin{center}
\includegraphics*[width=3.6cm,height=3.6cm]{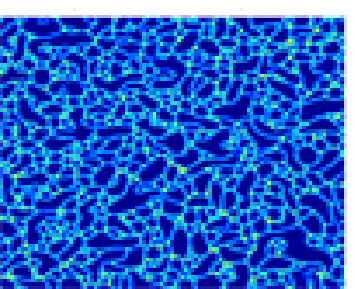}
\includegraphics*[width=3.6cm,height=3.6cm]{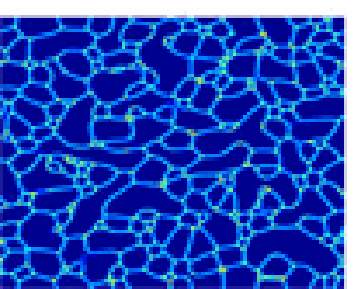}
\includegraphics*[width=3.6cm,height=3.6cm]{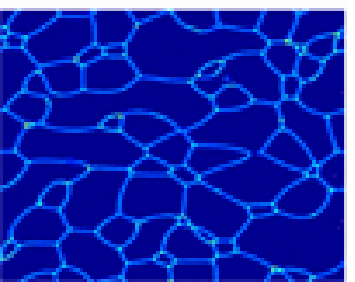}
\includegraphics*[width=3.6cm,height=3.6cm]{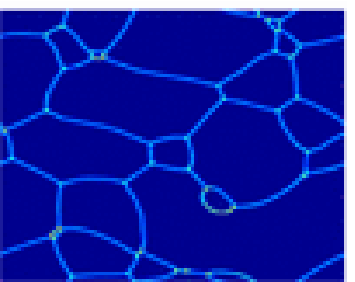}
\end{center}
\caption{
The matter-era evolution of the domain wall network for the ideal model with $N=7$. Note that only stable Y-type junctions survive. From left to right and top to bottom, the horizon is approximately 1/16, 1/8, 1/4 and 1/2 of the box size respectively.
}
\label{ideal7}
\end{figure}
%%%%%%%%%%%%%%%%%%%%%%%%%%%%%%%%%%%%%%%%%%%%%%%%%%%%%%%%

To confirm our description of the model (\ref{ideal}) we have simulated a number of realizations of the ideal class of models, that is the potential Eq. (\ref{ideal}) for different values of $N$ up to $N=10$. In all these simulations we have confirmed that only stable Y-type junctions are formed. In addition, any X-type junction imposed in the network by hand have been destroyed quickly. Two examples are shown in Figs. \ref{ideal4} and \ref{ideal7}, respectively for $N=4$ and $N=7$. Note in particular that these results confirm that in the Carter (pentahedral) model only Y-type junctions survive. It is very interesting to study several network properties, such as the degree of instability of the network (along the lines outlined above) or the degree of convergence towards a scaling solution \cite{SIMS1,SIMS2,AWALL} as a function of the number of fields $N$. This is again a task that is left for future study.

\section{\label{conc}Conclusions}

We have studied several typical domain wall forming models in two spatial dimensions, and discussed the conditions under which various types of defect junctions can exist. In particular, we focused on the BBL \cite{BAZEIA} and Kubotani \cite{KUBOTANI} models, both of which have been studied in the past. We have highlighted the situations where Y-type junctions are stable, those where only X-type ones are stable, and those where both types co-exist. Using both analytic and numerical techniques we have shown that topological and energetic arguments both contribute to the types and behavior of the allowed junctions, and we have identified their respective roles. A through understanding of the effects of both of these mechanisms is vital for a proper assessment of the cosmological consequences of defect models with junctions. This is particularly true for the case of cosmic (super)strings, which haven't as yet been explored in quantitative detail.

Our analysis also allowed us to distinguish which features of a particular model are crucial for its behavior and which ones are, to some extent at least, irrelevant details. The crucial step in the analysis is the simple but previously neglected result (which we have discussed in Paper I) that in models with more than two vacua, if all the domain walls connecting the various vacua have equal energies then only Y-type stable junctions can be stable. This in turn allowed us to define and explicitly construct a realization of what would be the ideal model in terms of would-be frustrated networks. Granted that this is a toy model, it does have the advantage of including all of the features that are desirable in terms of frustration (and none of the undesirable ones), as well as the further advantage of being simple enough to be easy to explore numerically. This model is of $\phi^6$ type, and our construction will work for any number of real scalar fields. as well as for any number of spatial dimensions. In addition, we have also provided a different realization, which is to some extent simpler (being of $\phi^4$ type) but arguably less general, since it applies only to an even number of scalar fields---though one could of course claim that it applies to any number of complex fields.

These results strongly support our no frustration conjecture. In all our 2D numerical simulations, including those of the ideal model, we find a dynamical behavior consistent with a convergence towards a scaling solution \cite{SIMS1,SIMS2,AWALL} and no evidence of any behavior that could eventually lead to frustration. (We also note that recent work by Battye \textit{et al.} provides further support for this.) We therefore conclude that in two spatial dimensions no frustrated network can be formed dynamically out of random initial conditions that would mimic a realistic phase transition. Notice that we are making no claims about the possibility of designing (by hand) a lattice-type configuration that could (in rather particular and/or idealized cases, such as the static limit) be stable against small perturbations. That is of course an interesting mathematical problem, but whether or not such a possibility exists is of very little consequence for the cosmological behavior of this type of defects. 

What now remains to be done is to extend this analysis to the more realistic case of three spatial dimensions. With our identification of the ideal class of models, this means are now available to undertake this analysis. Notice that the increased dimensionality will have non-trivial effects in the behavior of the junctions. Nevertheless, we have reasons to believe that the no frustration conjecture will also hold in three spatial dimensions. We leave a detailed analysis of this issue, as well as that several other interesting questions that we have highlighted at various points of this paper, for forthcoming work.

\section*{Acknowledgements}

This work was done in the context of the ESF COSLAB network and funded by FCT (Portugal), in the framework of the POCI2010 program, supported by FEDER. Specific funding came from grant POCI/CTE-AST/60808/2004 and from the Ph.D. grant SFRH/BD/4568/2001 (J.O.) J.M. and R.M. are supported by the Brazilian government (through CAPES-BRASIL), specifically through grants BEX-1970/02-0 and BEX-1090/05-4.

\bibliography{junctions}

\end{document}